\documentclass[fleqn,usenatbib]{mnras}
\usepackage[T1]{fontenc}
\DeclareRobustCommand{\VAN}[3]{#2}
\let\VANthebibliography\thebibliography
\def\thebibliography{\DeclareRobustCommand{\VAN}[3]{##3}\VANthebibliography}
\usepackage{amsmath, amssymb, graphicx, amsfonts, CJK, upgreek,ulem}
\title{Quakes of Compact Stars}
\author[Ruipeng Lu et al.]{
Ruipeng Lu,$^{1}$
Han Yue,$^{1}$\thanks{yue.han@pku.edu.cn}
Xiaoyu Lai,$^{2}$
Weihua Wang,$^{3}$
Shenjian Zhang$^{4}$ and
Renxin Xu$^{5,6}$\thanks{r.x.xu@pku.edu.cn}
\\
$^{1}$School of Earth and Space Sciences, Peking University, Beijing 100871, China\\
$^{2}$Department of Physics and Astronomy, Hubei University of Education, Wuhan 430205, China\\
$^{3}$College of Mathematics and Physics, Wenzhou University, Wenzhou 325035, China\\
$^{4}$Department of Earth and Space Sciences, Southern University of Science and Technology, Shenzhen 518055, China\\
$^{5}$School of Physics and State Key Laboratory of Nuclear Physics and Technology, Peking University, Beijing China, 100871\\
$^{6}$Kavli Institute for Astronomy and Astrophysics, Peking University, Beijing 100871, China
}
\date{Accepted XXX. Received YYY; in original form ZZZ}
\pubyear{2022}
\begin{document}
\label{firstpage}
\pagerange{\pageref{firstpage}--\pageref{lastpage}}
\maketitle
\begin{abstract}
Glitches are commonly observed for pulsars, which are explained by various mechanisms. One hypothesis attributes the glitch effect to the instantaneous moment of inertia change of the whole star caused by a starquake, which is similar to earthquakes caused by fast dislocation occurring on planar faults for the static stress, though the quake-induced dynamics responsible for glitch (superfluid vortex vs. pure starquake) remains still unknown. However, a theoretical model to quantitatively explain the stress loading, types of starquakes, and co-seismic change of moment of inertia is rarely discussed. In this study, we incorporate elastic deformation theories of earthquakes into the starquake problems. We compute the field of stress loading associated with rotation deceleration and determine the optimal type of starquakes at various locations. Two types of pulsar structure models, i.e. neutron and strangeon star models, are included in the computation and their differences are notable. Our calculation shows that the observed glitch amplitude can be explained by the starquakes in the strangeon star model, though the required scaled starquake magnitude is much larger than that occurred on the Earth. We further discuss the possibility to compute the energy budget and other glitch phenomena using the starquake model in the elastic medium framework.
\end{abstract}

\section{Introduction} \label{sec1}

What's the nature of bulk strong matter (i.e., ``gigantic nucleus'') in the Universe?
This is a question meaningful not only in thought experiments but also in understanding realistic stuff of compact stars and even dark matter, to be solved potentially in the era of gravitational-wave astronomy~\citep[e.g.,][]{Baiotti2019}.
Historically, exactly 90 years ago, Lev Landau proposed that neutron (``protons and electrons ... very close together'' in his words) matter forms when a huge number of atomic nuclei come in close contact after a gravitational collapse~\citep{Landau1932}, but the building units of such resultant matter could be quarks or strangeons if the symmetry restoration of three-flavour quarks ($u$, $d$ and $s$) works there (see, e.g., \cite{xlx2021} for a brief note about the long history).
How can we test these speculated models by observations?
It is popularly thought that pulsar glitch behavior would be an effective probe into the interior structure of a compact star, and this topic is then the focus of this work.

\begin{figure*}
\includegraphics[width=\textwidth]{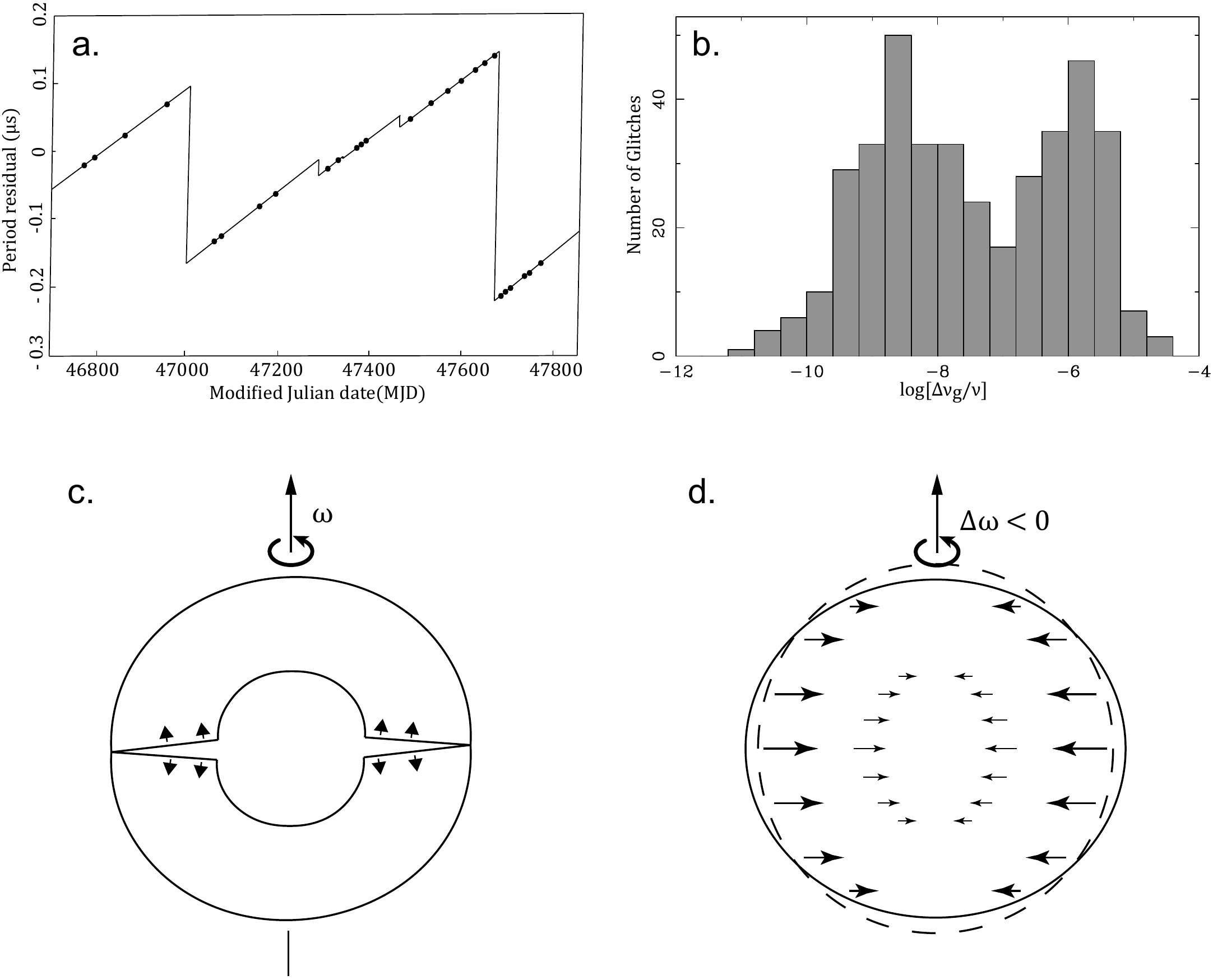}
\caption{(a) Temporal variation of the rotation period of PSR 1737-30 is plotted as black dots after removing a reference period of 606.59181ms. Linear extrapolation of gradual and sudden rotation change is denoted as black curves in which five glitch phenomena are observed~\citep{McKenna1990PSR173730}. (b) The number vs. magnitude (logarithm fraction of $\Delta\nu_g/ \nu$ ) histogram of all glitch phenomena occurred from 1990 to 2011 is plotted. The majority of glitch phenomena lie between $10^{-10}\sim10^{-5}$ ~\citep{Yu2012Detection}.(c) A classical starquake model in \protect\cite{Baym1971Neutron} is plotted, in which deceleration of rotation causes an opening in the neutron star crust, thus instantaneously reducing its rotation inertia and increasing rotation velocity.(d) The deformation of a spinning-down pulsar.Spinning velocity reduction introduces centripetal deformation and stress loading of pulsars.}
\label{f1}
\end{figure*}

As the final stage of a massive star, a pulsar is a fast-spinning (commonly with periods from a few milliseconds to a few seconds) compact object, which radiates electromagnetic waves from its magnetic poles. Such periodic signals can be observed on the Earth, which gradually decrease with time as the rotational energy is reduced by ejected pair-plasma and electromagnetic radiation. Glitch phenomena (the sudden change of spinning velocity) are occasionally observed for pulsars, which provide invaluable information on their internal structures (Figure \ref{f1} a$\&$b). There are several candidate mechanisms to explain the glitch phenomena, such as magnetospheric instabilities, pulsar disturbance by a planet, hydrodynamic instabilities, starquakes and vortex unpinning. Among them, the starquake (Figure \ref{f1} c$\&$d) and vortex unpinning models are the most widely accepted interpretations~\citep{Chamel2008Physics}.

The starquake model resembles the elastic-rebound mechanism of earthquakes, which is driven by a gradual stress loading from the decreased rotation~\citep{Baym1971Neutron,Franco2000Quaking}. Once such stress loading exceeds the failing criterion, an earthquake is excited to release the loaded stress. As a consequence, the elastic budge, which is formed at a higher rotation velocity is reduced and so is the total moment of inertia. The whole sequence occurs in a short time, which is equivalent to the elastic wave travel time over the pulsar (within $\sim 10$ ms $\sim r/ \sqrt{\mu/\rho^2}$, which is short compared with the glitch's unresolvable spin-up time and several days relaxation time \citep{Yu2012Detection}, with pulsar radius $r$, shear modulus $\mu$ and density $\rho$), thus causing an instantaneous increase in spinning velocity.

The pulsar quake behaves differently for a neutron star (NS) or a strangeon star (SS).
NS model progresses for a long time, with some open challenges in either micro-physics or astrophysics~\citep{Burgio2021}, whereas large-amplitude glitches remained difficult to explain  \citep{Crawford2003Comparison}.
Attention should be called that pulsars would in fact be strange quark stars if the hypothesis that strange quark matter could be more stable than nuclear matter is correct~\citep{Blaschke2005,Weber2005}.
Nevertheless, in view of different manifestations of pulsar-like compact stars, an SS model was proposed by~\citet{Xu2003Solid} (``Strangeon'' is coined by combining ``strange nucleon'', see~\cite{Xu:2016uod} and \cite{Lai:2017ney} for the details).
Preliminary investigations of pulsar glitches are presented in an SS
model~\citep{ Zhou:2004ue,Peng:2007eq,Zhou:2014tba}, and detailed modeling confronted with observations has also striven: \cite{Lai2018Pulsar} showed the potential of starquake models of SSs to explain the correlation between the recovery coefficient and relative glitch amplitude, while \cite{Wang:2020xsm} discussed the glitch activity of normal radio pulsars.
It is worth noting that, for the first repeating fast radio burst (FRB), the frequency vs. magnitude relationship of quakes/glitches follows a power-law relationship similar to earthquakes \citep{Wang2018FRB,Howitt2018Nonparametric}, namely the Gutenberg-Ritcher relationship \citep{1944BuSSA..34..185G}, and quake-induced magnetospheric activity could be responsible for the production of the bulk of energetic bunches of repeating FRBs in order to understand both observational features of the time-frequency drifting and the polarization~\citep{WangWY2022a,WangWY2022b}.
Recently, the global parameters~\citep{Gao:2021uus} of rotating SSs (mass, radius, moment of inertia,tidal deformability, quadrupole moments, and shape parameters) and the oscillation modes~\citep{LiHB2022} of non-rotating SSs have also been investigated
in full general relativity.
Therefore, quantitative analysis of starquakes under NS and SS models will be helpful to distinguish these two models and draw further implications to the inner properties of pulsars.

Additionally, in a conventional NS starquake model like which proposed by \cite{Baym1971Neutron} (Figure \ref{f1}c), an opening occurs in the inner wall of the crust near the equator. However, such quakes with opening mechanisms are rarely observed on the Earth, because of the presence of litho-static stresses. The litho-static stress increase with depth, which is typically orders of magnitude larger than the tectonic stress at the seismogenic depths. Several works show that the fracture under compressive stress is a shear failure instead of an opening failure under tensile stress \citep{HOEK2014287,2005PhRvL..94i4301Z}. Therefore shear failure is always preferred for earthquake models, which is rarely discussed for starquakes also with static stress. Meanwhile, a theoretical framework remains to be built to explain the location, magnitude and types of starquakes, and to predict the change of moment of inertia in a starquake . It will allow us to make quantitative comparisons and predictions of starquake models and to draw conclusions with higher precision. Most previous works focus on how stars' evolutions trigger starquakes~\citep{2022MNRAS.511.3365G,2022MNRAS.514.1628K}. A framework to explain the influence of starquakes on stars' evolution is needed. In this paper, we do calculations on the change of the inertia produced by a starquake, whose results would be useful in further research about relaxing processes or stress accumulation, for both neutron stars and strangeon stars.

On the other hand, related frameworks have been built for earthquake models under elastic theory \citep{Dahlen1971Excitation,Dahlen1973Correction,Chao1987Changes}.
Related theories have been adopted to investigate the earthquake-induced rotational velocity change of the Earth, though the theoretical prediction (less than $10{\rm \upmu s}$ such as \cite{2006SGeo...27..615G} and \cite {2013TAOS...24..649X}) is close to the observational precision (from International Earth Rotation and Reference Systems Service with the error more than $10{\rm \upmu s}$ \citep{2011Metro..48S.165G}) of the period of earthquake rotation (length of a day). Related elastic theories can be directly adopted for starquakes.

\subsection{Stress development to produce frequently large glitches in starquake model}

It is a general concept that, due to the low mass of crust with low elasticity, in conventional starquake models of pulsar glitches the required stress develops too slowly to produce large glitches as often as they are observed ~\citep{Pines1974}, as explained below.
Before the occurrence of the next quake, the star must develop an excess oblateness $\Delta\epsilon$ which has been released in the last quake.
The time interval between two successive glitches $t_{\rm q}$ (from this glitch to the next) of a pulsar (with mass $M$, radius $R$, spin frequency $\omega$ and the spin-down rate $\dot\omega$) could be derived to be~\citep{Baym1971Neutron,Lai2018Pulsar}
\begin{equation}
    t_{\rm q}=\frac{2A(A+B)}{BI_0}\frac{|\Delta\epsilon|}{\omega\dot\omega},\label{tq1}
\end{equation}
where $A=3GM^2/25R$, $B=\mu V_{\rm c}$ ($V_{\rm c}$ and $\mu$ are respectively the volume and the mean shear modulus of the solid crust (for an NS) and the whole star (for an SS)) and $I_0$ is the moment of inertia of the star's crust (for an NS) and whole star (for an SS) without spin.
Our computation is under the assumption that the star is idealized as a two-layers sphere for an NS and a uniform elastic sphere for an SS.

If the sudden spin-up of a glitch is attributed to $\Delta\epsilon$, then the glitch magnitude $\Delta\omega/\omega=|\Delta\epsilon|$.
For the Crab pulsar whose typical glitch magnitude is $\Delta\omega/\omega\sim 10^{-9}$, the observed values of $t_{\rm q}$ can be consistent with that predicted in Eq.(\ref{tq1}).
However, for the Vela pulsar whose typical glitch magnitude is $\Delta\omega/\omega\sim 10^{-6}$, the neutron star model indicates that even if the star is entirely solid, $t_{\rm q}$ derived in Eq.(\ref{tq1}) is much larger than the observed values.
This Vela-discrepancy seems to mark the failure of the starquake mechanism for an explanation of glitches.

As a matter of fact, starquakes manifested in the form of glitches are different in the solid strangeon star model, being discussed in the previous paper by~\cite{Lai2018Pulsar}.
Our results have shown consistency with the observed values, as explained in the following.
Motivated by the observational fact that glitches with small amplitudes recover almost completely, but those with large amplitudes recover negligible, we introduce an $\eta$-value to reflect the plastic flow (un-recoverable) triggered by oblateness development characterized by the $\epsilon$-value.
The inner motion of the star during a starquake is not only the change of oblateness ($\Delta\epsilon$), but also a redistribution of matter (parameterized as $\Delta\eta$), both of which would change the moment of inertia $I$ of the star.
The glitch magnitude is then written as $\Delta\omega/\omega=|\Delta\epsilon|+|\Delta\eta|$.
With the assumption that only $\Delta\epsilon$, not $\Delta\eta$, would
lead to a release of stress during glitches, then we can see that $\Delta\epsilon$ is not directly related to $\Delta\omega/\omega$, which means that we cannot predict the time interval $t_{\rm q}$ in Equation(\ref{tq1}) only from $\Delta\omega/\omega$.
It is consequently reasonable that, the Crab pulsar and the Vela pulsar have nearly the same values of $t_{\rm q}$ although their glitch magnitudes differ by at most three orders of magnitude.
Our results have shown consistency with the observed values, as shown in Fig.4 of~\cite{Lai2018Pulsar}.

Another concerned aspect about starquakes in strangeon stars is the energy release problem.
Previous works found that, starquakes in the strangeon star will be accompanied by huge gravitational energy~\citep{Zhou:2014tba} or strain energy release due to its large shear modulus~\citep{Wang:2020xsm}.
Meanwhile, young pulsars (such as the Crab pulsar) are thought to release more strain energy based on their large predicted oblateness and reference oblateness.
However, radio~\citep{2018MNRAS.478.3832S} and X-ray observations~\citep{2020A&A...633A..57V} on the Crab pulsar following its largest 2017-glitch revealed no flux enhancement.
To sum up, no flux enhancement in the radio or high energy band has been found following glitches in radio pulsars at present (except the high magnetic field pulsars PSR J1119-6127 and PSR J1846-0258).

The detailed mechanisms that take energy away remain unknown, but we think at least three mechanisms could be responsible.
Firstly, most glitches are accompanied by increases in spin-down torques~\footnote{\url{http://www.jb.man.ac.uk/pulsar/glitches/gTable.html}~\citep{2011MNRAS.414.1679E},
and \url{http://www.atnf.csiro.au/people/pulsar/psrcat/glitchTbl.html}.}, which corresponds to an increase in the rotational energy loss rate, i.e., the wind braking~~\citep{TongH2013}.
Starquakes may lead to energetic particle flow or magnetic reconnection, producing spin-down enhancement and radiative changes as observed in the Crab pulsar~\citep{2020NatAs...4..511F} and the Vela pulsar \citep{2018Natur.556..219P}, et al.,
part of this energy could also be injected into the pulsar wind nebula as observed by Ge et al.~\citep{2019NatAs...3.1122G}.
Secondly, starquakes may excite some oscillation modes or asymmetry in neutron star structure, inducing short-timescale gravitational waves~\citep{2015MNRAS.446..865K}, gravitational wave burst~\citep{2020JApA...41...14L}, and/or transient gravitational waves~\citep{2020MNRAS.498.1826G,2020MNRAS.498.3138Y}.
Thirdly, a small part of the energy release could also be dissipated during the post-glitch relaxation process in the form of heat energy. For the compact start glitch effects, post-glitch recovery/adjustment is commonly observed~\citep{1992Natur.359..706L}, which is associated with the non-elastic response of the compact star medium. For earthquakes, this effect is commonly recognized as post-seismic relaxation, which is associated with slow after-slips on the fault plane and viscous relaxation in the lower-curst or upper mantle. In both mechanisms, elastic energy is dissipated as heat energy~\citep{2006GeoJI.167..397P}.

One might also suspect the possibility that the compact stars reach the breaking condition and suffer starquakes frequently without external influences such as accretion~\citep{2018arXiv180404952F}, especially when both the shear modulus and the crustal breaking strain angle is high~\citep{Horowitz2009BreakingSO,Baiko2018Breaking,Wang:2020xsm}.
Theoretically, this problem is hard to answer from first principle calculations, due to the complexity of rotational evolution history and the uncertainty of the internal structure of pulsar-like compact stars.
One possibility may be, as pointed out by~\cite{Giliberti2019Modelling}, the crust may never fully relax after a glitch. In other words, stars always stay around the critical point where starquakes occur.
Observationally, signatures of starquakes have been observed in normal radio pulsars~\citep{2020NatAs...4..511F}, in high magnetic field pulsars such as PSR J1119-6127~\citep{2015MNRAS.449..933A}, and also in magnetars such as SGR J1830-0645~\citep{2022ApJ...924L..27Y}, et al.

From the discussions above, it is evident that one cannot reproduce Vela-like glitches with the pure starquake mechanism in conventional NS models, but this does not mean that a quantitative calculation of the stress loading in NSs is not necessary.
Even in the popular scenario of superfluidity, the vortex unpinning might be triggered by a starquake, whereas quake-induced magnetic reconnection could power magnetar's giant flares or bursts.
In this sense, a detailed investigation of starquakes would be welcome, in both NS and SS models, in order to find possible observational evidence for either of the models.
In this work, we adopt the self-gravitated spherically symmetric elastic body theory to discuss the elastic stress distribution of a pulsar model caused by centrifugal body force reduction during the spinning-down, and the associated preferred starquake types under this theory. We also discuss the moment of inertia change produced by a starquake and compare the observed glitch amplitude with scaled earthquake magnitude. The efficiency of the moment of inertia change in starquakes at different locations is also discussed. Our calculations are derived for NS and SS models respectively, drawing respective implications for starquakes occurring in either condition. It needs to be pointed out that for pulsars with extremely dense matters, a strict derivation should be made under the general relativity framework, while as the first step, the elastic theory draws a first-order approximation for the starquake problems.

\vspace{6mm}

\section{Methods} \label{sec2}
\subsection{Strain accumulation produced by spin down}

\cite{Backus1967Converting} describes a method to calculate displacement fields introduced by body forces for a spherically symmetric gravitational model. We computed the field of strain tensors caused by spin-down and determine the quake mechanisms at different locations. Then we computed the inertia change by quake dislocations.

\begin{figure}
\includegraphics[width=\columnwidth]{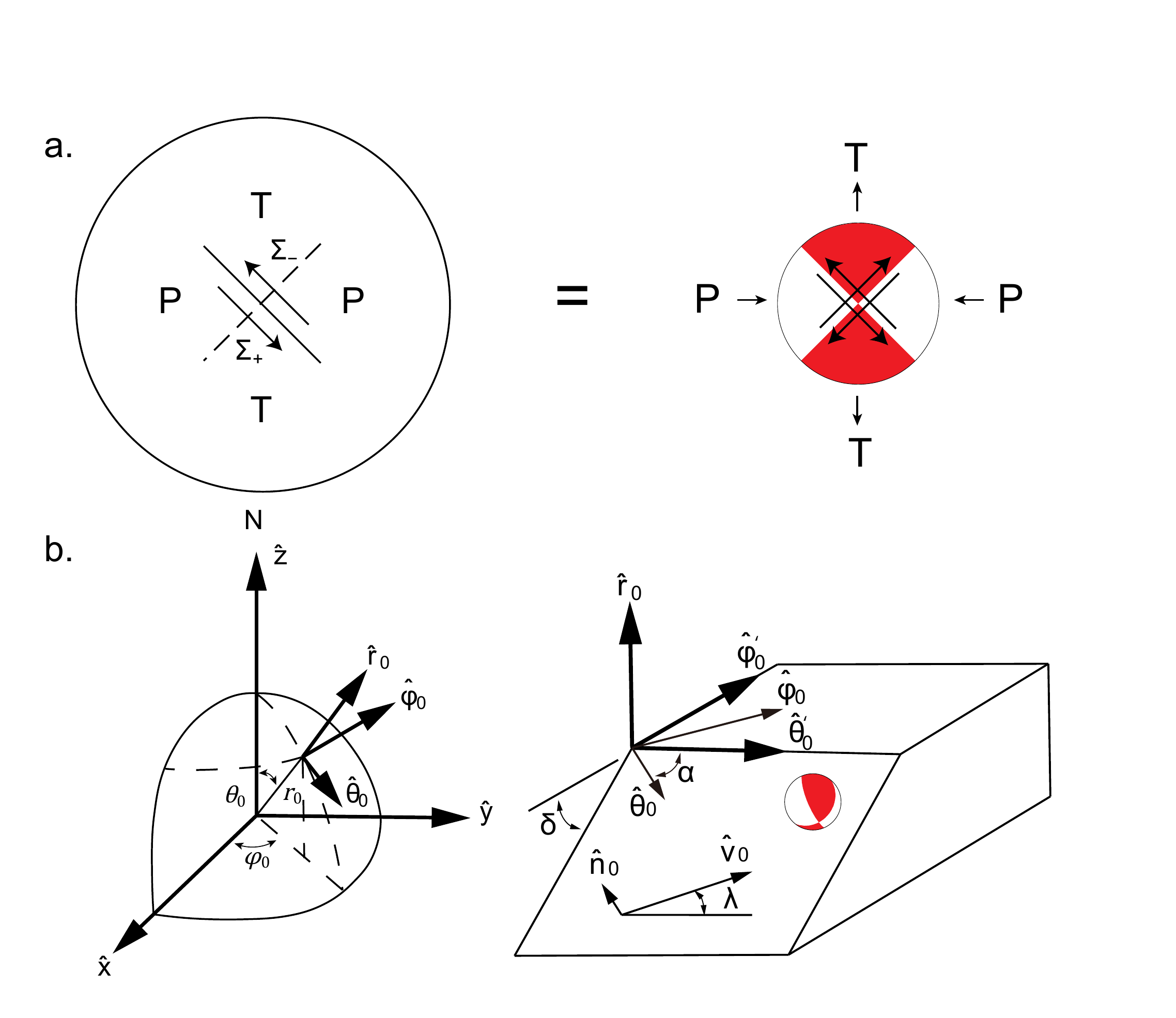}
\caption{(a) A diagram of the seismic representation theory. The displacement field caused by a dislocation on a fault surface $\Sigma$ (left) is equivalent to that caused by a double force couple (right). A double shear force couple is commonly presented by compressional and dilatational force couples, that is visualized by a focal mechanism, in which the compressional (P) and dilatational (T) axis are denoted as white and red domains. (b) The symbol convention in this article. $r_0,\theta_0,\varphi_0$ are the spherical coordinates of the fault location.
$\hat{r}_0,\hat{\theta}_0,\hat{\varphi}_0$ are local coordinates, which are related to vertical, south and east directions. $\alpha,\delta,\lambda$ define the geometry of a focal mechanism in a local coordinate. Slips along the direction $\hat{v}_0$ on a fault plane (with a normal direction of $\hat{n}_0$) are denoted as a focal mechanism.}
\label{f2}
\end{figure}

The centrifugal body force reduction caused by spinning deceleration  is expressed by
\begin{equation}
\boldsymbol{f}(r, \theta, \varphi)=\Delta \omega^{2} \rho r\left(\sin ^{2} \theta \hat{\boldsymbol{r}}+\sin \theta \cos \theta \hat{\boldsymbol{\theta}}\right).
\label{1}
\end{equation}
in which the definition of spherical coordinates is illustrated in  Figure \ref{f2}. In summary, the displacement potential function can be described by a Poisson equation, whose solution is the linear summation of an infinite series of spherical harmonics that
\begin{equation}
Y_{l}^{m}(\theta, \varphi)=(-1)^{m}\left[\frac{2 l+1}{4 \pi}\right]^{\frac{1}{2}} \cdot\left[\frac{(l-m) !}{(l+m) !}\right]^{\frac{1}{2}} P_{l}^{m}(\cos \theta) \exp (i m \varphi) .
\end{equation}
Because of the special form of centrifugal forces, free boundary condition at the star surface and displacement continuous condition in the star center, only the zeroth- and second-order spherical harmonics have non-zero coefficients. The coefficients of spherical harmonics are numerically calculated by radial integration over the presumed star elastic model, which is further used to calculate the displacement and strain field.  All calculation details are summarized in the appendix in Equation \eqref{12}-\eqref{*20}. For the NS model, we use a two layers model \citep{Giliberti2019Incompressible}. The core is assumed to be liquid with zero shear modulus.  For the SS model, we use a solid sphere model \citep{Xu2003Solid}. Details of the elastic properties of the NS and SS models are summarized in Table \ref{t1}. Both of them are incompressible models. We use such a setting because the bulk modulus of a compact star is much larger than the shear modulus \citep{Chamel2008Physics} and compressibility will not affect our conclusions. This peculiarity could still keep for denser matter of strangeon, at supranuclear density. The associated displacement field is described by
\begin{equation}
\mathbf{u}=\sqrt{\frac{5}{16\pi}} (\mathcal{U}_{2}^{0}(r)(3\cos ^{2} \theta -1)\hat{\mathbf{r}} -3\mathcal{V}_{2}^{0}(r)\sin 2\theta\hat{\boldsymbol{\theta}}).
\label{*2}
\end{equation}

\begin{table*}
\centering
\caption{Parameters of pulsar models}
\begin{tabular}{lccccr}
		\hline
		Type & Core radius & Core density & Crust thickness & Crust density & Shear modulus\\
		\hline
        NS & $9.5\times10^{5}$cm &  $6.6\times10^{14}{\rm gcm^{-3}}$ & $5\times10^{4}$cm & $6.6\times10^{13}{\rm gcm^{-3}}$ & $10^{30}{\rm dyncm^{-2}} $ \\
		   SS & $10^{6}$cm & $6.6\times10^{14}{\rm gcm^{-3}}$ & - & - & $10^{35}{\rm dyncm^{-2}}$\\
        \hline
\label{t1}
	\end{tabular}
\end{table*}
Earthquakes are driven by stress loading related to the strain field with Hooke's law. Under spherical coordinates, elastic strains can be calculated by Equation \eqref{*2}, which is realized by numerical derivatives of the spherical harmonic coefficients
\begin{equation}
\begin{aligned}
\varepsilon_{rr}&=\sqrt{\frac{5}{16\pi}} \frac{d\mathcal{U}_{2}^{0}}{d r}(3\cos ^{2} \theta -1) \\
\varepsilon_{\theta \theta}&=\sqrt{\frac{5}{16\pi}} (\mathcal{U}_{2}^{0}(3\cos ^{2} \theta -1) -3\mathcal{V}_{2}^{0}\sin 2\theta)r^{-1} \\
\varepsilon_{\varphi \varphi}&=\sqrt{\frac{5}{16\pi}} (\mathcal{U}_{2}^{0}(3\cos ^{2} \theta -1) -3\mathcal{V}_{2}^{0}\cos^2 \theta)r^{-1} \\
\varepsilon_{r \theta}&=\sqrt{\frac{45}{64\pi}} (\frac{\mathcal{V}_{2}^{0}}{r}-\frac{d\mathcal{V}_{2}^{0}}{d r}-\frac{\mathcal{U}_{2}^{0}}{r})\sin2\theta\\
\varepsilon_{r \varphi}&=\varepsilon_{\theta \varphi}=0.
\label{*3}
\end{aligned}
\end{equation}

\begin{figure*}
\includegraphics[width=\textwidth]{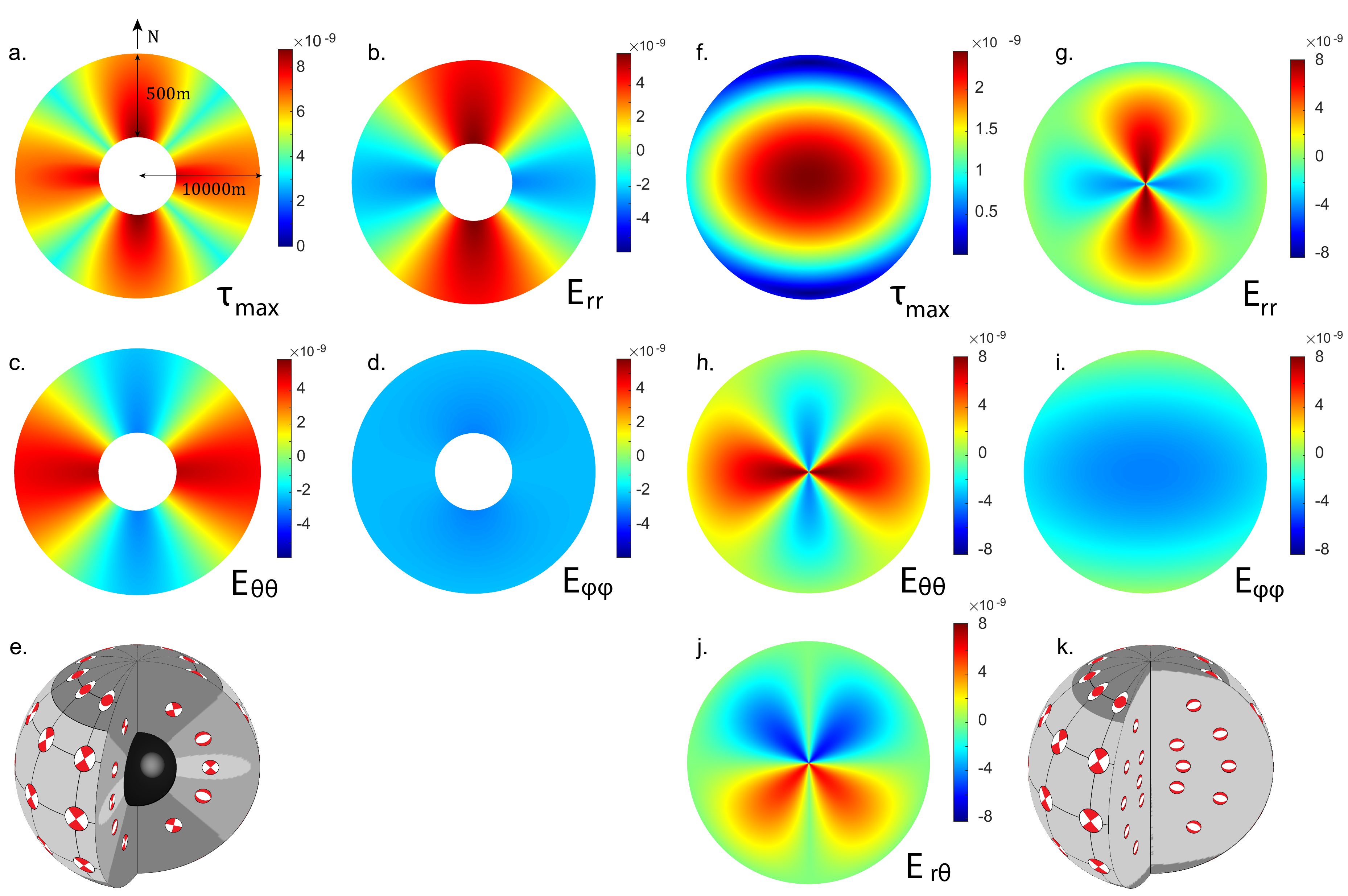}
\caption{Plots of the strain distribution on a cross-section through the center of a pulsar with $\Delta\omega^2=-1{\rm rad^2 s^{-2}}$. Strain solutions of NS and SS are plotted in the left and right panels, respectively. (a-d) The maximum shear strain and each strain component (i.e. $E_{rr}, E_{\theta\theta}, E_{\varphi\varphi}$) are plotted in each panel. Note the crust of the neutron is plotted on an exaggerated scale. (e) The focal mechanisms of optimal faulting are plotted on the surface and cross-sections of NS in a 3D perspective. (f-j) the same as (a-e) for the maximum shear strain and each strain component (i.e. $E_{rr}, E_{\theta\theta}, E_{\varphi\varphi}, E_{r\theta}$), respectively. Note $E_{r\theta}$ is none zero for SS. (k) is the same as (e) for optimal focal mechanisms on SS. }
\label{f3}
\end{figure*}

Based on the superposition principle, for the distribution of strain loading caused by rotation deceleration, its value is proportional to $\Delta\omega^2$. To visualize the spinning down induced stress loading, we assume $\Delta\omega^2=-1{\rm rad^2 s^{-2}}$ as a representative to calculate and plot the results in Figure \ref{f3} that observation shows $\Delta\omega^2=10^{-4}\sim 10^{-1 }{\rm rad^2 s^{-2}}$ between two glitches \citep{Espinoza2011ASO}. We plot each component of strain fields and focal mechanism (the meaning is given later and shown in Figure \ref{f2}a) in Figure \ref{f3} for NS and SS models, respectively. Because the displacement field is rotationally symmetric, the strain field is identical in any profiles cutting the rotational axis, we only plot the strain field over a profile that cut the pulsar along the rotation axis. It is worth noting that earthquakes are driven by shear stress, which is commonly related to the second invariant (maximum shear strain) of all strain components, thus we plot the maximum shear strain in Figure \ref{f3}a$\&$f for NS and SS models, respectively, which reflects locations most likely to be ruptured by starquakes in both models. For the NS model, the equator and pole areas of the crust have a similar amount of deviatoric stress building. Because the liquid core has zero shear strength, no shear strain is built in the core, thus all starquakes should occur in the thin crust. For the SS model, shear stress accumulation is most significant in the center, while shear stress in the crust is smaller than that in the center. It is also noted that shear stresses near the equator are much larger than that near the poles. Such a contrast (one order of magnitude difference) is more significant than that between the surface and interior shear stresses near the equator (about two times difference).

Although the maximum shear strain field tells the location of starquakes, their rupture types need to be presented by focal mechanisms. As presented by the earthquake source representation theorem: the displacement field generated by a spatially compact fault surface is equivalent to that produced by a double-force couple (double couple model). The double couple model is commonly presented by a focal mechanism and visualized by a four-lobed "beach ball". A focal mechanism uses two nodal planes (one of which is the fault plane) to cut the sphere into four equal domains. The center axis of the white and red domains are the compressional (P) and dilatational (T) axis, which are 45°from the slip vector (Figure \ref{f2}a). The cross line of the two nodal planes is the N axis. Depending on the plunge angle of the three axes, the types of earthquakes are classified as thrust events (T axis close to vertical), normal events (P axis close to vertical ) and strike-slip events (N axis close to vertical), which reflects compressional, dilatational and shear strain release in the horizontal direction, respectively.

Considering the relative amplitude of deviatoric strain components, the starquakes that occurred in different parts of the pulsar surface and interior are classified into different types (Figure \ref{f3}e$\&$k). Different types of starquakes have different values of $\Delta I/I$, because the deviatoric strain components will change the seismic moment tensor $M$ under the same scalar moment $M_0$, as shown in Equation\eqref{**7} \eqref{4} \eqref{5} .
In addition, the types of starquakes determine the calculations of both the change of $I$ in starquakes (Figure \ref{f7}) and the post-seismic adjustment.

For the NS model, both equator and pole areas in the crust have large shear stress loading. For its equator area, the inner and outer surfaces of the crust are characterized by normal and strike-slip events respectively. The pole areas of the NS crust are characterized by thrust faulting. For the SS model, the majority of the star volume is characterized by strike-slip faulting, except for the surfaces of pole areas, which are characterized by thrust faulting.

\subsection{Change of moment of inertia caused by a starquake}
After calculating the likely location and types of starquakes, we need to calculate the moment inertia change produced by a starquake, which is helpful to understand the relationship between the starquake magnitude and observed rotation velocity change. The equivalent body force of a tangential displacement dislocation is a double couple force expressed in \cite{Dahlen1973Correction}
\begin{equation}
\boldsymbol{e}(r, \theta, \varphi)=-\mathbf M\cdot\nabla \delta\left(\boldsymbol r-\boldsymbol{r_{0}}\right)
\label{2}
\end{equation}
where $\boldsymbol r_0$ is the position vector of the fault and $\mathbf M$ is the seismic moment tensor defined as
\begin{equation}
\mathbf M=M_0(\hat{\mathbf{n}}_{0}\hat{\mathbf{v}}_{0}+\hat{\mathbf{v}}_{0}\hat{\mathbf{n}}_{0}); M_0=\mu A d
\label{**7}
\end{equation}
where $M_0$ is the scalar seismic moment, $\mu$ is the shear modulus, $A$ is the rupture area, $d$ is the rupture displacement, $\hat{\mathbf{n}}_{0}$ is its normal direction, $\hat{\mathbf{v}}_{0}$ is the strike direction, can be expressed in terms of the strike $\alpha$, dip $\delta$, and slip $\lambda$ (Figure \ref{f2}b) in
\begin{equation}
\begin{aligned}
&\hat{\mathbf{n}}_{0}=\cos \delta \mathbf{r}_{0}+\sin \alpha \sin \delta \hat{\boldsymbol{\theta}}_{0}-\cos \alpha \sin \delta \hat{\boldsymbol{\varphi}}_{0} \\
&\hat{\mathbf{v}}_{0}=\sin \delta \sin \lambda \hat{\mathbf{r}}_{0}+(\cos \alpha \cos \lambda-\sin \alpha \cos \delta \sin \lambda) \hat{\boldsymbol{\theta}}_{0} \\
&+(\sin \alpha \cos \lambda+\cos \alpha \cos \delta \sin \lambda) \hat{\boldsymbol{\varphi}}_{0} \\
&\hat{\mathbf{n}}_{0} \cdot \hat{\mathbf{v}}_{0}=0.
\end{aligned}
\label{3}
\end{equation}
After getting the displacement field, the inertia change caused by dislocation can be expressed in the form
\begin{equation}
\begin{aligned}
\Delta I=\Gamma{M_{0}}&=\frac{4}{3}\Gamma_{0} M_{rr}+2\Gamma_{1} M_{\theta\varphi}\cos ^{2} \phi \\
&\qquad+\frac{2}{3}\Gamma_{2}M_{rr}(1-3\sin ^{2} \phi) +\Gamma_{3}M_{r\varphi} \sin 2 \phi
\label{4}\\
&=M_0(\Gamma_{0} j_0+\Gamma_{1}j_1+\Gamma_{2}j_2+\Gamma_{3}j_3)
\end{aligned}
\end{equation}
where $\Delta I$ is the moment of inertia change cause by a starquake with seismic moment tensor of $\mathbf M$ and $M_{rr}, M_{\theta\varphi}, M_{r\varphi}$ are the corresponding components in spherical coordinates. Note that $\phi$ is the latitude, different from $\varphi$ which means the azimuth.
$M_0$ has a dimension of torque (unit of dyncm) and $\Gamma$ has a dimension of inertia dividing torque (unit of ${\rm s^2}$). $\Gamma_i$ are functions related to the star model and fault depths, which is calculated by numerical integration of model elastic parameters shown in Equation\eqref{18}-\eqref{*30}. $j$ functions are defined as:
\begin{equation}
\begin{aligned}
j_{0} & =\frac{4}{3} \sin 2 \delta \sin \lambda \\
j_{1} & =-(2 \sin 2 \alpha \sin \delta \cos \lambda+\cos 2 \alpha \sin 2 \delta \sin \lambda) \cos ^{2} \phi \\
j_{2} & =\frac{2}{3} \sin 2 \delta \sin \lambda(1-3\sin ^{2} \phi) \\
j_{3} & =(\cos \alpha \cos \delta \cos \lambda-\sin \alpha \cos 2 \delta \sin \lambda) \sin 2 \phi
\end{aligned}
\label{5}
\end{equation}
which depend on the angular fault parameters and starquake latitude, while they are independent of the star model and the fault depth. Amplitudes of glitch events can be calculated by angular momentum conservation, as \citep{Baym1971Neutron}
\begin{equation}
\frac{\Delta \nu}{\nu}=-\frac{\Delta I}{I},
\label{6}
\end{equation}
where $\nu$ is the frequency of a pulsar's rotation, equalled with the pulse frequency people received on the Earth.
Relationships defined in Equation \eqref{4}-\eqref{6} provide a quantitative relation between the starquake moment $M_0$ and glitch magnitude ${\Delta \nu}/{\nu}$, which are used to evaluate the magnitude of possible starquakes.

\section{RESULTS and discussion} \label{sec:results}
\subsection{Order of magnitude of \texorpdfstring{$\Gamma$ }  .functions }

As shown in equation 8, the $\Gamma$ functions are the scales between the seismic moment and moment of inertia changes, thus their amplitudes are critical to estimating the relationship between starquake magnitude and the glitch amplitude. We compute $\Gamma$ functions of two models (Table \ref{t1}) and plot their depth distribution in Figure \ref{f4}.
$\Gamma_0$ is related to the compressibility of stars (see Appendix \ref{appendixB}), so both models have zero value of $\Gamma_0$ due to incompressibility.

For the NS model, $\Gamma_1, \Gamma_2$ are $10^{-8}{\rm s^2}$ and $\Gamma_3$ is nearly zero, with about a 20 per cent change over depth in the crust. This variation shows the depth variation of NS starquakes produces an insignificant impact on the moment of inertia change. For the SS model, $\Gamma_1, \Gamma_2, \Gamma_3$ are in the order of $10^{-9} {\rm s^2}$, which is one order of magnitude smaller than that of the NS model. Especially, at shallow depths ($<$ 500m), the order of magnitude of $\Gamma$ functions is $10^{-10} {\rm s^2}$, which is two orders of magnitude smaller than that in the NS model. This difference means starquakes in the SS model are less efficient in changing the moment of inertia of the whole pulsar. Also different from the NS model, the amplitude of $\Gamma$ functions in the SS model varies by about one order of magnitude for star quakes occurring near the surface and near the center. Considering overburden pressures increase with depth, which increases the normal stress and frictional stress on fault planes, starquakes may be more likely to occur near the surface. We infer the typical $\Gamma$ function magnitude is more presented by that near the surface, thus the typical $\Gamma$ function magnitudes are estimated as
\begin{equation}
\Gamma_{NS}\sim10^{-9} {\rm s^2};\Gamma_{SS}\sim10^{-10} {\rm s^2}
\label{7}
\end{equation}
for two typical starquake models, respectively.
\begin{figure*}
\includegraphics[width=\textwidth]{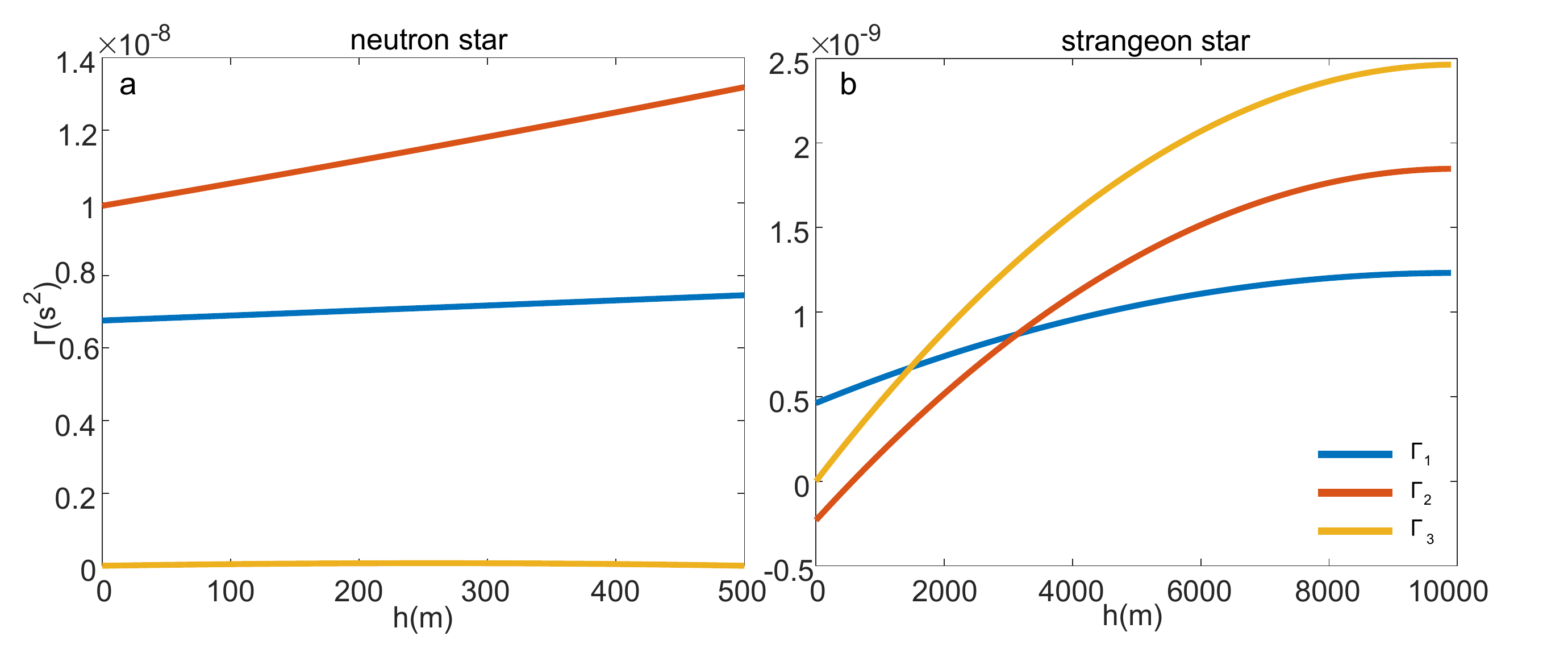}
\caption{The depth distributions of integration factors ($\Gamma_1, \Gamma_2, \Gamma_3$) are plotted in blue, red and yellow curves, respectively. Solutions of NS and SS are plotted in a and b panels, respectively.$\Gamma_0$ is not plotted because it is always zero. }
\label{f4}
\end{figure*}

\subsection{The estimation of the seismic moment of a starquake\label {4.1}}

\begin{figure*}
\includegraphics[width=\textwidth]{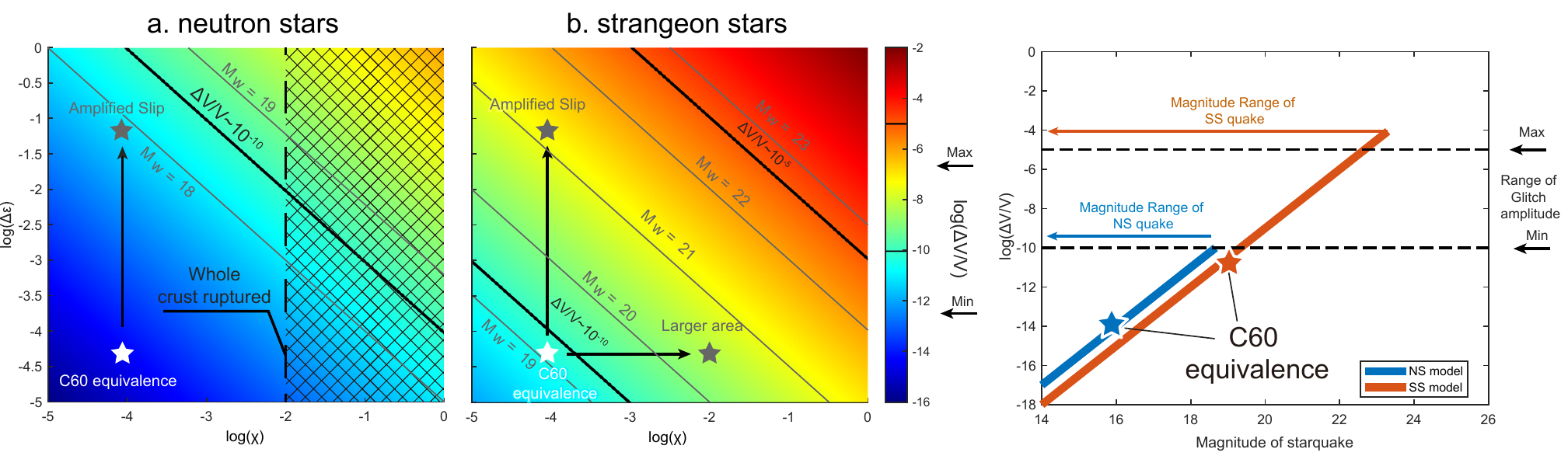}
\caption{Dependence of glitch magnitude (log ($\Delta\nu/\nu$)) on the star quake strain drop ($\varepsilon$) and moment ratio ($\chi$). Glitch magnitude of NS and SS are plotted in the left and right panels, respectively. The maximum and minimum glitch magnitude ($10^{-10}\sim10^{-5}$) are bounded by black lines. The line of $10^{-5}$ is not given in NS for it is over the theoretical maximum. The shadow in (a) mark the area that $\chi$ is not allowed in NS. A scaled earthquake of the largest observed earthquake (i.e. 1960 Chile earthquake) is respectively marked as white stars in both figures as references. Grey lines are isolines of starquake magnitude with corresponding $\Delta\varepsilon$ and $\chi$ in NS and SS models, respectively. (c) shows the comparison between NS and SS models. The range of observation glitch amplitude is shown as horizontal dot lines. C60 in the two models is respectively marked as the blue and orange star.}
\label{f5}
\end{figure*}

It is noted the starquake induced moment of inertia change is proportional to its seismic moment, the scalar amplitude of which is presented in Equation\eqref{**7}. As little is known about the physical properties of starquakes, including their area or displacement, we estimated the scale of starquakes on pulsars by dimensional analysis.
We assume the starquake ruptures on a circle fault plane, whose dimension is $r$. The fault area is thus $r^2$. Distance to be influenced by the rupture is also $r$, yielding a characteristic volume to be influenced by the starquake as $r^3$. We define a dimensionless factor of $\chi$, which is the ratio between the starquake characteristic volume and the volume of the star. For the order of magnitude purpose, we neglect the difference between radius and diameter.
\begin{equation}
\chi=\frac{r^3}{R^3}
\label{9}
\end{equation}
which defines the starquake moment vs. conceptual moment release if the whole pulsar is ruptured. Substitute Equation\eqref{**7}, the scalar seismic moment can be expressed in
\begin{equation}
M_0=\mu\Delta\varepsilon r^3
\end{equation}
where $\Delta\varepsilon=d/r$ is the strain drop of a starquake.
As a starquake scale is definitely smaller than the scale of the pulsar, $\chi $ is less than 1. For earthquakes, the value is less than $10^{-4}$, whose upper limit is defined by the largest observed earthquake. i.e. 1960 Chile earthquake, with $M_0=2\times10^{30}{\rm dyncm}$, and typical $\mu\sim10^{11}{\rm dyncm^{-2}}, R=6371{\rm km}, \Delta\varepsilon\sim10^{-4}$ for averaged earth model \citep{Dziewonski1981Preliminary}. With $\chi$ and Equation\eqref{4}\eqref{6}, glitch value can be expressed in
\begin{equation}
\frac{\Delta\nu}{\nu}=\frac{\chi\Delta\varepsilon\mu R^3\Gamma}{I}.
\label{10}
\end{equation}
Various starquake strain drops are reported \citep{Horowitz2009BreakingSO,Baiko2018Breaking,Ruderman1991Neutron}, which lie in the range of $10^{-5}\sim10^{-1}$. We use typical $\Gamma$ values of NS and SS models, i.e. ($10^{-9}$ and $10^{-10}$), to plot the trade-off relationship between the moment of inertia change $\Delta\nu/\nu$, in the starquake strain drop $\Delta\varepsilon$ and factor $\chi$ domains in Figure \ref{f5}.

In comparison with the observations, the starquake model appears to be a possible explanation of glitch phenomena. To give a common sense about the amplitude of starquakes, we make the dimensional and magnitude scale of the largest earthquake (1960 Mw = 9.5 Chile earthquake) in this diagram (namely C60 equivalent event).
A pulsar is much denser and more uniform than the Earth, so a quake occurring on a pulsar is supposed to rupture a larger relative magnitude and present a higher strain drop. The comparison of quakes on pulsars with the ever-known largest earthquake could give us a sense of the magnitudes and strain drops of  pulsars' quakes.
When scaling the C60 event, we maintain the dimension ratio between the quake and star/earth diameter (the $\chi$ factor) and strain drop $\Delta\varepsilon$, which is realized by  proportionally reducing the earthquake scale and slip when shrinking the earth radius to the pulsar radius. While the seismic moment is scaled up to $10^{40}$dyncm and $10^{45}$dyncm due to the shear modulus increase, which is equivalent to an $Mw=16$ and $Mw = 19$ starquake for the NS and SS models, respectively.

For the SS model, the C60 equivalent event is at the lower end of the observed glitch amplitude (Figure \ref{f5}b), while for the NS model, the C60 equivalent event can not produce the observed glitch amplitude (Figure \ref{f5}a). To generate starquakes of larger magnitude, either strain drops or the scales of starquakes need to be increased, to shift the starquake magnitude to the top-right. For pulsar models, both trends are possible. Reported strain drops for star quakes are between $10^{-5}\sim10^{-1}$ which is $3\sim4$ orders of magnitude higher than the earthquake, which presents a vertical shift of the C60 equivalent event. on the other hand, earthquakes are constrained by the dimension of Earth's plate boundaries, which can barely grow to a scale comparable to the Earth's radius. However, for pulsar models, if the medium is more homogeneous and globally subject to a uniform stress loading, a starquake may grow to a  scale comparable to the star radius, yielding $\chi$ value to be $0.01\sim0.1$. When applying such stress drops and scale ratio to the C60 equivalence in the SS model, the observed glitch magnitude can be produced by $M_w=20\sim23$ starquakes. However, for the NS model, since starquakes are constrained within the crust, the maximum value of $\chi$ is 0.01. If both maximum strain drop of 0.1 and maximum $\chi$ are considered, which produces a $M_w\sim$19 starquake, the seismic moment can barely meet the lower end of the observed glitch amplitude $\Delta\nu/\nu\sim10^{-10}$. Thus the NS model can not generate a strong enough quake to produce larger glitches. From this perspective, the SS starquake model is more plausible to explain the observed glitch phenomenon.

\subsection{The effect of elastic property}

Our computation is under the assumption that the star is idealized as a two-layers sphere for an NS and a uniform elastic sphere for an SS. In fact, pulsars have a multiple-layers structure like the Earth \citep{Giliberti2019Modelling}, the elastic property we choose to represent the total star will produce a difference. So it is necessary to discuss the effect of shear modulus uncertainties.

\begin{figure*}
\includegraphics[width=\textwidth]{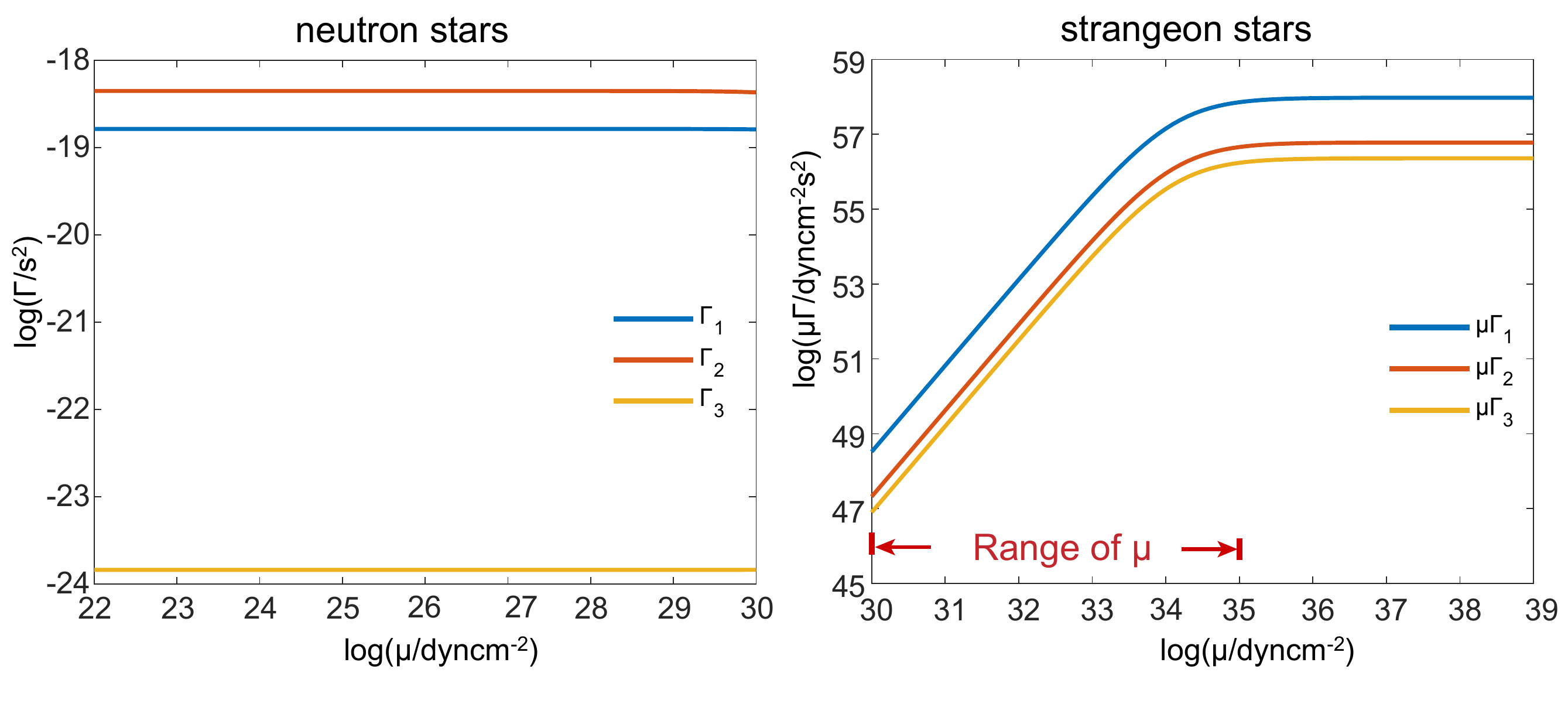}
\caption{(a) The relationships between $\Gamma$ and $\mu$ of NS and SS are plotted in each panel. For the two models, a standard crust thickness of $h=200$m is used for calculation. Estimated ranges of reasonable shear modulus ($10^{30}\sim10^{35}{\rm dyncm^{-2}}$) are marked.}
\label{f6}
\end{figure*}

We simplify the Equation\eqref{10}, that
\begin{equation}
\frac{\Delta \nu}{\nu} \propto \frac{\Gamma\mu \Delta \varepsilon}{\rho}
\label{11}
\end{equation}
where strain drop $\Delta \varepsilon$ is independent of other parameters. The change of shear modulus $\mu$ will influence $\Gamma$; the change of density $\rho$ will influence $\Gamma$ and $I$. The equation of state determined the relation between $\mu$ and $\rho$.
For NS, there are several types of research about the equation of state and $\mu\propto\rho$ is a proper estimate \citep{Chamel2008Physics}. Due to the lack of direct experimental evidence, there is, unfortunately, no certain expression to relate $\rho$ and $\mu$ for strangeon matter now~\citep{Xu2003Solid}.
Nevertheless, it's a good approximation to have a uniform density for a $\sim 1.4M_\odot$ bare strange star due to sharp surface~\citep{Alcock1986}.
This approximation applies better for strangeon stars with stiffer equation of state.
Therefore, we use constant $\mu$ in our calculations, but changing with a large order of magnitude ($\mu=10^{30}\sim10^{35}{\rm dyncm^{-2}}$), as shown in the right plot of Fig.~\ref{f6}.
Then the derived relationship is $\Delta \nu/\nu \propto \Gamma$ for NS and $\Delta \nu/\nu \propto \Gamma\mu$ for SS. The effects of changing $\mu$ for the NS and SS models are plotted in Figure \ref{f6}, respectively. It shows that for the NS model, the effect of changing $\mu$ is not significant, which means the relationship between starquake moment and glitch magnitude relationship derived in Equation \eqref{7} holds up for a broad range of elastic properties. While, for the SS, the effect of changing $\mu$ is significant. The relationship derived in Equation \eqref{7} uses $\mu=10^{35}{\rm dyncm^{-2}}$, which lies at the higher end of the reasonable range of shear modulus. Adopting a lower shear modulus produces a proportional effect on the associated glitch amplitude.

\subsection{The effect of different failing criteria}
\begin{figure}
\includegraphics[width=\columnwidth]{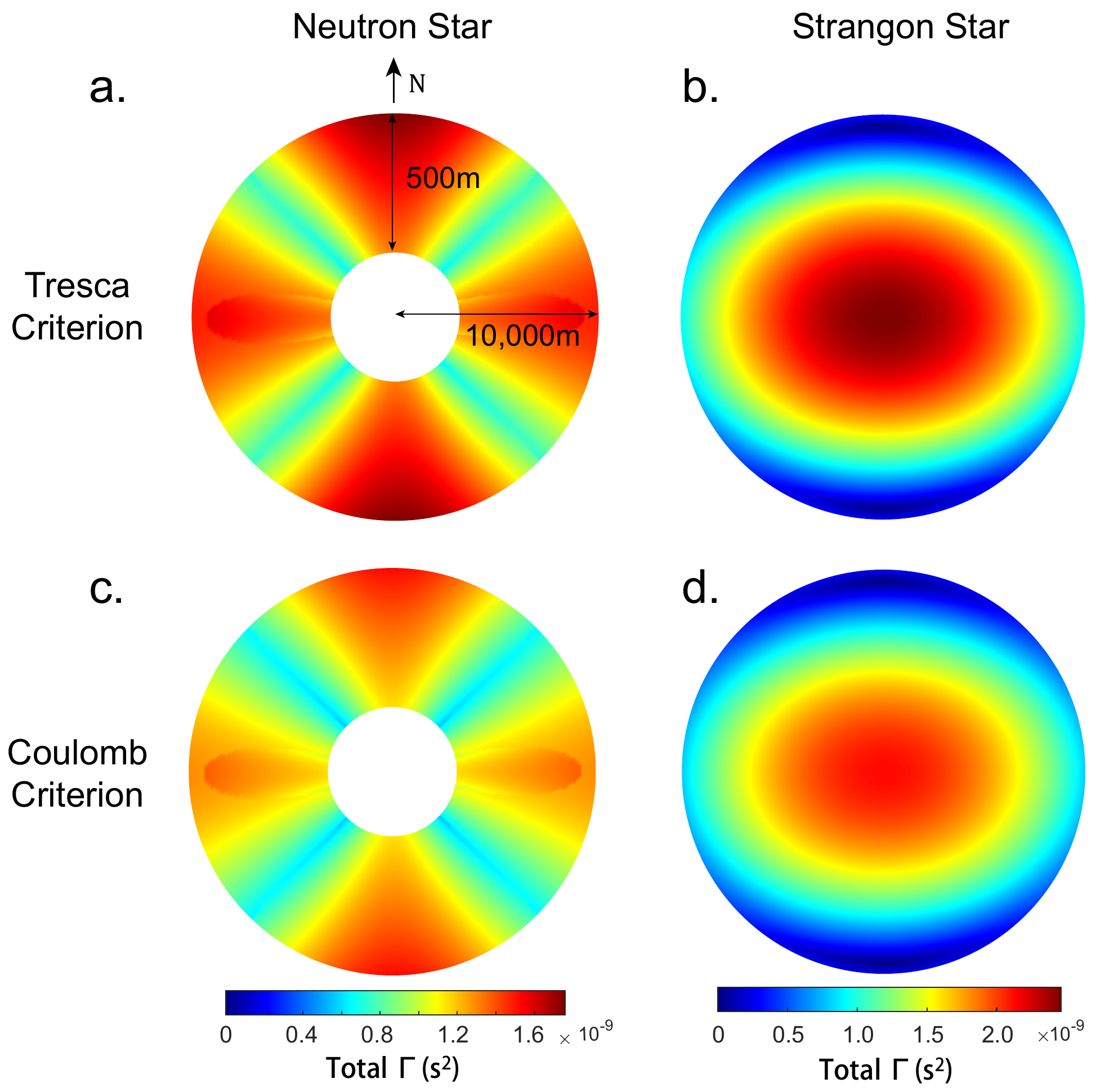}
\caption{$\Gamma$ values calculated by optimal faults are images at the respective fault locations on the cross-sections of pulse stars. NS and SS are plotted in the left and right panels, respectively. Focal mechanisms calculated using Tresca and Coulomb criteria are plotted in the top and bottom panels, respectively. }
\label{f7}
\end{figure}

In the computation through of this study, we followed \cite{Giliberti2019Incompressible} and use the Tresca criterion to determine the crust breaking, which assumes the rupture occurs along the maximum shear direction (45° from the maximum compressional stress).  However, the Coulomb criterion is more commonly adopted for earthquakes, which considers both normal and shear stress on a fault plane, thus a rupture occurs when the shear stress is larger than the frictional stresses (normal stress times frictional parameter).  For the Coulomb criterion and common frictional parameter of 0.6, the optimal fault directions are 30° from the maximum compressional stress instead of 45° from the maximum compressional stress direction as that predicted by the Tresca criterion.  We compute the optima focal mechanisms at each point of the NS and SS models following two criteria and compute the associated $\Gamma$ function amplitudes for both NS and SS models, respectively (Figure \ref{f7}). The comparison shows that the $\Gamma$ function pattern is similar to the maximum shear strain produced by spinning down,  which satisfies an empirical judgment that starquakes occurred at higher shear stress concentration and are more efficient in changing the pulsar moment of inertia.  Different failing criteria produce slight differences in the $\Gamma$ function amplitude, while for the order of magnitude estimated realized in this work, such differences can be neglected.

\section{Conclusion}
In this work, we derived a framework to calculate the starquake-related elastic strain theory. We adopt an existing earthquake theory, which calculates elastic stress loading on a self-gravitated elastic layered symmetric sphere to solve starquake problems. Our calculation shows:
\begin{itemize}
\item[1.]
Strain loading and focal mechanisms vary significantly on different pulsar models, i.e. neutron star and strangeon star models.
\item[2.]
Starquakes in the strangeon model are more plausible to explain the observed glitch amplitude, while starquakes in the NS model are not strong enough to explain the observation.
\item[3.]
Our calculation is robust for the uncertainties of NS elastic parameters, while it is sensitive to the SS elastic parameters. Variations in the failing criterion do not influence the order of magnitude estimation significantly.
\end{itemize}

This work presents a first-order approximation for the starquake problems, which can be used to estimate other starquake-related problems including the stress loading rate, energy budget and relaxation mechanisms. For example, strain loading can be analytically calculated using Equation \eqref{*3} to estimate strain loading caused by spinning deceleration between two glitches. The accumulated strain can be compared with the starquake magnitude derived by Equation \eqref{4} from the following glitch amplitude, which draws light to the energy budget and material strength of the starquake cycles.
\section*{Acknowledgements}

This work was supported by the National SKA Program of China (2020SKA0120100,2020SKA0120300), the National Natural Science
Foundation of China (Grant No. 42174059) and the Young Top-notch Talent Cultivation Program of Hubei Province.

\section*{ORCID iDs}
Lai, X. Y. https://orcid.org/0000-0002-3093-8476
\\
Wang, W. H. https://orcid.org/0000-0003-1473-5713
\\
Xu, R. X. https://orcid.org/0000-0002-9042-3044

\section*{Data availability}
The data underlying this article are available at:
\url{https://github.com/Millaboreus80/QCSsupport.git}.

\bibliography{ref}
\bibliographystyle{aasjournal}

\appendix
\label{appendix}
\section{Displacement field calculation}
For a spherical, non-rotating, elastic, isotropic star model, the displacement motivated by a body force $\mathbf{f}$ can be calculated by solving the equations \citep{Backus1967Converting}
\begin{equation}
\begin{aligned}
&-\rho_{0} \nabla \phi_{1}-\rho_{1} \nabla \phi_{0}-\nabla\left(\mathbf{v} \cdot \rho_{0} \nabla \phi_{0}\right)+\nabla \cdot \mathbf{E}+\mathbf{f}=0 \\
&\nabla^{2} \phi_{1}=4 \pi G \rho_{1} \\
&\rho_{1}=-\nabla \cdot\left(\rho_{0} \mathbf{v}\right) \\
&\mathbf{E}=\lambda(\nabla \cdot \mathbf{v}) \mathbf{I}+2 \mu\left[\nabla \mathbf{v}+(\nabla \mathbf{v})^{\mathrm{T}}\right]
\end{aligned}
\label{12}
\end{equation}
where $\rho_0$, $\phi_0$ is the density and gravitational potential without applying body force, $\rho_1$,$\phi_1$ is the disturbance brought by the body force, $\mu$ is the shear modulus and $\lambda$ is the bulk modulus which is infinity in incompressible models. The equations include Newton's second law, gravity law, continuity equation, and constitutive equation. Solving the equations will get the displacement expression.

These equations may be converted into a set of scalar equations by using spherical harmonic expansion
\begin{equation}
\begin{aligned}
\mathbf{v}(\mathbf{r})&=\hat{\mathbf{r}} \mathcal{U}(\mathbf{r})+\nabla_{1} \mathcal{V}(\mathbf{r})-\hat{\mathbf{r}} \times \nabla_{1} \mathcal{W}(\mathbf{r}) \\
\mathbf{f}(\mathbf{r})&=\hat{\mathbf{r}} \mathcal{A}(\mathbf{r})+\nabla_{1} \mathcal{B}(\mathbf{r})-\hat{\mathbf{r}} \times \nabla_{1} \mathcal{C}(\mathbf{r})
\end{aligned}
\label{13}
\end{equation}
where
\begin{equation}
\begin{aligned}
\nabla_{1} &=\hat{\boldsymbol{\theta}} \frac{\partial}{\partial \theta}+\hat{\boldsymbol\varphi} \frac{1}{\sin \theta} \frac{\partial}{\partial \varphi} \\
\mathcal{U}(\mathbf{r}) &=\sum_{l=0}^{\infty} \sum_{m=-l}^{l} \mathcal{U}_{l}^{m}(r) Y_{l}^{m}(\theta,\varphi) \\
\mathcal{V}(\mathbf{r}) &=\sum_{l=0}^{\infty} \sum_{m=-l}^{l} \mathcal{V}_{l}^{m}(r) Y_{l}^{m}(\theta,\varphi) \\
\mathcal{W}(\mathbf{r}) &=\sum_{l=0}^{\infty} \sum_{m=-l}^{l} \mathcal{W}_{l}^{m}(r) Y_{l}^{m}(\theta,\varphi) \\
\mathcal{A}(\mathbf{r}) &=\sum_{l=0}^{\infty} \sum_{m=-l}^{l} \mathcal{A}_{l}^{m}(r) Y_{l}^{m}(\theta,\varphi) \\
\mathcal{B}(\mathbf{r}) &=\sum_{l=0}^{\infty} \sum_{m=-l}^{l} \mathcal{B}_{l}^{m}(r) Y_{l}^{m}(\theta,\varphi) \\
\mathcal{C}(\mathbf{r}) &=\sum_{l=0}^{\infty} \sum_{m=-l}^{l} \mathcal{C}_{l}^{m}(r) Y_{l}^{m}(\theta,\varphi),
\end{aligned}
\label{14}
\end{equation}
and
\begin{equation}
Y_{l}^{m}(\theta, \varphi)=(-1)^{m}\left[\frac{2 l+1}{4 \pi}\right]^{\frac{1}{2}}\left[\frac{(l-m) !}{(l+m) !}\right]^{\frac{1}{2}} P_{l}^{m}(\cos \theta) \exp (i m \varphi).
\label{*17}
\end{equation}
For the orthogonality of spherical harmonics, substitute Equation\eqref{13}\eqref{14}, Equation\eqref{12} can be reduced for given $l\neq0$ and $m$ to
\begin{equation}
\label{15}
\begin{aligned}
\frac{d}{d r} \mathcal{U}_{l}^{m}&=-\lambda(r \beta)^{-1}[2 \mathcal{U}_{l}^{m}-l(l+1) \mathcal{V}_{l}^{m}]+\beta^{-1} P \\
\frac{d}{d r} \mathcal{V}_{l}^{m}&=r^{-1}[\mathcal{V}_{l}^{m}-\mathcal{U}_{l}^{m}]+\mu^{-1} Q \\
\frac{d}{d r} \mathcal{W}_{l}^{m}&=r^{-1}\mathcal{W}_{l}^{m}+\mu^{-1} R \\
r \frac{d}{d r} P&=\left(4 r^{-1} \gamma-4 \rho_{0} g_{0}\right) \mathcal{U}_{l}^{m}-l(l+1)\left(2 r^{-1} \gamma-\rho_{0} g_{0}\right) \mathcal{V}_{l}^{m}\\
&\qquad+2\left(\lambda \beta^{-1}-1\right) P+l(l+1) Q+r \rho_{0} g_{1}-\mathcal{A}_{l}^{m} r \\
r \frac{d}{d r} Q&=\left(\rho_{0} g_{0}-2 r^{-1} \gamma\right) \mathcal{U}_{l}^{m}-r^{-1}[2 \mu-l(l+1)(\gamma+\mu)] \mathcal{V}_{l}^{m}\\
&\qquad-\lambda \beta^{-1} P-3 Q+\rho_{0} \phi_{1}-\mathcal{B}_{l}^{m} r \\
r \frac{d}{d r} R&=r^{-1} \mu\left(l(l+1)-2\right)\mathcal{W}_{l}^{m}-3 R-\mathcal{C}_{l}^{m} r\\
\frac{d}{d r} \phi_{1}&=g_{1}-4 \pi G \rho_{0} \mathcal{U}_{l}^{m} \\
\frac{d}{d r} g_{1}&=-2 r^{-1} g_{1}+l(l+1) r^{-2} \phi_{1}+l(l+1)4 \pi G \rho_{0} r^{-1} \mathcal{V}_{l}^{m}. \\
\end{aligned}
\end{equation}
When $l=0$, $\mathcal{V}_{0}^{0}, \mathcal{W}_{0}^{0}$ vanish after derivation, Equation\eqref{12} can be reduced to
\begin{equation}
\label{16}
\begin{aligned}
\frac{d}{d r} \mathcal{U}_{0}^{0}&=-2\lambda(r \beta)^{-1}\mathcal{U}_{0}^{0}+\beta^{-1} P \\
r \frac{d}{d r} P&=\left(4 r^{-1} \gamma-4 \rho_{0} g_{0}\right) \mathcal{U}_{0}^{0}+2\left(\lambda \beta^{-1}-1\right) P+r \rho_{0} g_{1}-\mathcal{A}_{0}^{0} r \\
\frac{d}{d r} \phi_{1}&=g_{1}-4 \pi G \rho_{0} \mathcal{U}_{0}^{0} \\
\frac{d}{d r} g_{1}&=-2 r^{-1} g_{1}\\
\end{aligned}
\end{equation}
where
\begin{equation}
\label{17}
\beta=\lambda+2 \mu, \gamma=\lambda+\mu-\lambda^{2} \beta^{-1},
\end{equation}
$P,Q,R,\phi_1,g_1$ are auxiliary parameters, related to stress and gravity. The boundary conditions are $\mathcal{U}_{l}^{m},\mathcal{V}_{l}^{m}, \mathcal{W}_{l}^{m},P,Q,R,\phi_1,g_1$ must be continuous at anywhere, be finite at sphere centre, and $P=Q=R=0, l(l+1)\phi_1+g_1a=0$ at outside surface. Then we can get the arithmetic solution of the displacement field motivated by any body force.

Substitute Equation\eqref{*17} into the centrifugal body force and write in the form of Equation\eqref{13}, we have
\begin{equation}
\begin{aligned}
&\mathcal{A}_{0}^{0}=\frac{\sqrt{4\pi} \Delta \omega^2\rho r}{3} \\
&\mathcal{A}_{2}^{0}=-\frac{\sqrt{4\pi} \Delta \omega^2\rho r}{3\sqrt{5}} \\
&\mathcal{B}_{2}^{0}=-\frac{\sqrt{4\pi} \Delta \omega^2\rho r}{3\sqrt{5}}
\label{*20}
\end{aligned}
\end{equation}
and other $\mathcal{A}_{l}^{m}=\mathcal{B}_{l}^{m}=\mathcal{C}_{l}^{m}=0$. With the incompressible model shown in Table \ref{t1}, we got the numerical solution of $\mathcal{U}_{2}^{0}, \mathcal{V}_{2}^{0}$ and other displacement components are zero. Then we got the displacement and strain field expressed in Equation\eqref{*2}\eqref{*3}

\section{Inertia change calculation}

\label{appendixB}

For a known displacement field $\mathbf{v}$, the change of inertia can be expressed in the integral over the total space
\begin{equation}
\Delta I=\int \Delta \rho r^{2} \sin ^{2} \theta d V
\label{18}
\end{equation}
where
\begin{equation}
\Delta \rho=-\nabla \cdot(\rho \mathbf{v}),
\label{19}
\end{equation}
$\rho$ is the unperturbed density without any displacement. Using the integral over the total space instead of the sphere space avoids the confusion that using different treatments at the inner and outside surfaces to solve the discontinuity of density.

Substitute Equation\eqref{13}\eqref{14} into Equation\eqref{19} , the harmonic expansion of the density is
\begin{equation}
\begin{aligned}
&\Delta \rho=\sum_{l=0}^{\infty} \sum_{m=-l}^{l}\Delta  \rho_{l}^{m}(r) Y_{l}^{m}(\theta,\varphi), \\
&\Delta \rho_{l}^{m}=-\left(\partial_{r}\left(\rho \mathcal{U}_{l}^{m}\right)+\rho \frac{2 \mathcal{U}_{l}^{m}}{r}\right)+\rho \frac{l(l+1) \mathcal{V}_{l}^{m}}{r}.
\end{aligned}
\label{20}
\end{equation}

Substitute Equation\eqref{20} into Equation\eqref{18}, integral angle and radius respectively
\begin{equation}
\begin{aligned}
\Delta I&=\sum_{l=0}^{\infty} \sum_{m=-l}^{l}\int_0^{\infty} \Delta \rho_{l}^{m} r^{4}dr \int_{4\pi}  Y_{l}^{m}\sin ^{2} \theta d\omega\\
&=\sum_{l=0}^{\infty} \sum_{m=-l}^{l}\int_0^{\infty} \Delta \rho_{l}^{m} r^{4}dr \int_{4\pi}  \frac{2}{3} Y_{l}^{m}(\sqrt{4 \pi}Y_{0}^{0}-\sqrt{\frac{4 \pi}{5}}Y_{2}^{0}) d\omega\\
&=\frac{2}{3} \sqrt{4 \pi}\int_0^{\infty} \Delta \rho_{0}^{0} r^{4}dr-\frac{2}{3} \sqrt{\frac{4 \pi}{5}}\int_0^{\infty} \Delta \rho_{2}^{0} r^{4}dr.
\end{aligned}
\label{21}
\end{equation}

In this study, the body force is the equivalent body force of a tangential displacement dislocation shown in Equation\eqref{2}, its harmonic expansion can divide into three components $\left[\Lambda_{l}^{m}\right]_{1}, \left[\Lambda_{l}^{m}\right]_{2}, \left[\Lambda_{l}^{m}\right]_{3}$
\begin{equation}
\begin{aligned}
\mathcal{A}_{l}^{m}(r)&=M_{0}\left\{\frac{1}{r^{3}} \delta\left(r-r_{0}\right)\left[\left[\Lambda_{l}^{m}\right]_{3}-\frac{3}{2}\left[\Lambda_{l}^{m}\right]_{2}\right]\right.\\
&\left. \qquad-\frac{d}{d r}\left[\frac{1}{r^{2}} \delta\left(r-r_{0}\right)\right]\left[\frac{1}{2}\left[\Lambda_{l}^{m}\right]_{2}\right]\right\} \\
&=\mathcal{A}_{l,1}^{m}\left[\Lambda_{l}^{m}\right]_{1}+\mathcal{A}_{l,2}^{m}\left[\Lambda_{l}^{m}\right]_{2}+\mathcal{A}_{l,3}^{m}\left[\Lambda_{l}^{m}\right]_{3}, \\
l(l+1) \mathcal{B}_{l}^{m}(r)&=M_{0}\left\{\frac{1}{r^{3}} \delta\left(r-r_{0}\right)\left[\frac{1}{4} l(l+1)\left[\Lambda_{l}^{m}\right]_{2}+\left[\Lambda_{l}^{m}\right]_{1}\right.\right.\\
&\left.\left. \qquad-3\left[\Lambda_{l}^{m}\right]_{3}\right]-\frac{d}{d r}\left[\frac{1}{r^{2}} \delta\left(r-r_{0}\right)\right]\left[\Lambda_{l}^{m}\right]_{3}\right\}\\
&=l(l+1) (\mathcal{B}_{l,1}^{m}\left[\Lambda_{l}^{m}\right]_{1}+\mathcal{B}_{l,2}^{m}\left[\Lambda_{l}^{m}\right]_{2}+\mathcal{B}_{l,3}^{m}\left[\Lambda_{l}^{m}\right]_{3})\\
\mathcal{C}_{l}^{m}(r)&=0
\end{aligned}
\label{22}
\end{equation}
where
\begin{equation}
\begin{aligned}
{\left[\Lambda_{l}^{m}\right]_{1}=} &\left(n_{\theta} e_{\theta}-n_{\varphi} e_{\varphi}\right)\left[l(l+1) Y_{l}^{*}\left(\theta_{0}, \varphi_{0}\right)+2 \frac{\partial^{2}}{\partial \theta^{2}} \stackrel{*}{Y}_{l}^{m}\left(\theta_{0}, \varphi_{0}\right)\right] \\
&+2\left(n_{\theta} e_{\varphi}+n_{\varphi} e_{\theta}\right)\left[\frac{\partial}{\partial \theta}\left(\frac{1}{\sin \theta} \frac{\partial}{\partial \varphi} \stackrel{*}{Y}_{l}^{m}\left(\theta_{0}, \varphi_{0}\right)\right)\right] ;\\
{\left[\Lambda_{l}^{m}\right]_{2}=} & 4 n_{r} e_{r} \stackrel{*}{Y}_{l}^{m}\left(\theta_{0}, \varphi_{0}\right); \\
{\left[\Lambda_{l}^{m}\right]_{3}=} &\left(n_{r} e_{\theta}+n_{\theta} e_{r}\right) \frac{\partial}{\partial \theta} \stackrel{*}{Y}_{l}^{m}\left(\theta_{0}, \varphi_{0}\right) \\
&+\left(n_{r} e_{\varphi}+n_{\varphi} e_{r}\right)\left[\frac{\partial}{\partial \theta}\left(\frac{1}{\sin \theta} \frac{\partial}{\partial \varphi} \stackrel{*}{Y}_{l}^{m}\left(\theta_{0}, \varphi_{0}\right)\right)\right],
\end{aligned}
\label{23}
\end{equation}
$\theta_0,\varphi_0$ is the latitude and longitude of the fault.

Substitute Equation\eqref{22}\eqref{23} into Equation\eqref{20}, that
\begin{equation}
\begin{aligned}
&\Delta \rho_{l}^{m}=\Delta \rho_{l,1}^{m}\left[\Lambda_{l}^{m}\right]_{1}+\Delta \rho_{l,2}^{m}\left[\Lambda_{l}^{m}\right]_{2}+\Delta \rho_{l,3}^{m}\left[\Lambda_{l}^{m}\right]_{3},\\
&\Delta \rho_{l,i}^{m}=-\left(\partial_{r}\left(\rho \mathcal{U}_{l,i}^{m}\right)+\rho \frac{2 \mathcal{U}_{l,i}^{m}}{r}\right)+\rho \frac{l(l+1) \mathcal{V}_{l,i}^{m}}{r}
\end{aligned}
\label{24}
\end{equation}
where $\mathcal{U}_{l,i}^{m},\mathcal{V}_{l,i}^{m}$ is the numerial solution from substituting $\mathcal{A}_{l,i}^{m},\mathcal{B}_{l,i}^{m}$ into Equation\eqref{15}\eqref{16}.

Substitute Equation\eqref{24} into Equation\eqref{21}, that
\begin{equation}
\begin{aligned}
&\Delta I=M_0(\frac{2}{3} \sqrt{4 \pi}\Gamma_{0,1}^{0}\left[\Lambda_{0}^{0}\right]_{1}+
\frac{2}{3} \sqrt{4 \pi}\Gamma_{0,2}^{0}\left[\Lambda_{0}^{0}\right]_{2}+
\frac{2}{3} \sqrt{4 \pi}\Gamma_{0,3}^{0}\left[\Lambda_{0}^{0}\right]_{3} \\
&+\frac{2}{3} \sqrt{\frac{4 \pi}{5}}\Gamma_{2,1}^{0}\left[\Lambda_{2}^{0}\right]_{1}
+\frac{2}{3} \sqrt{\frac{4 \pi}{5}}\Gamma_{2,2}^{0}\left[\Lambda_{2}^{0}\right]_{2}+\frac{2}{3} \sqrt{\frac{4 \pi}{5}}\Gamma_{2,3}^{0}\left[\Lambda_{2}^{0}\right]_{3})
\end{aligned}
\end{equation}
where
\begin{equation}
\begin{aligned}
\Gamma_{l,i}^{m}=\int_0^{\infty}\Delta \rho_{l,i}^{m}r^{4}dr.
\label{*30}
\end{aligned}
\end{equation}
Notice that $\left[\Lambda_{0}^{0}\right]_{1}=\left[\Lambda_{0}^{0}\right]_{2}=0$. Briefly record $\Gamma_{0,1}^{0}, \Gamma_{0,1}^{2},\Gamma_{0,2}^{2}, \Gamma_{0,3}^{2}$ as $\Gamma_{0}, \Gamma_{1}, \Gamma_{2}, \Gamma_{3}$. Substitute Equation\eqref{3} into $\left[\Lambda_{0}^{0}\right]_{2}$, $\left[\Lambda_{2}^{0}\right]_{1}$, $\left[\Lambda_{2}^{0}\right]_{2}$, $\left[\Lambda_{2}^{0}\right]_{3}$ and record as $j_0, j_1, j_2, j_3$. We got Equation\eqref{4}\eqref{5}, where the significance of all symbols has been given.

\bsp
\label{lastpage}
\end{document}